\documentclass[reprint,superscriptaddress,amsmath,amssymb,aps,pra,floatfix]{revtex4-1}

\usepackage{amsmath}
\usepackage{graphicx}
\usepackage[usenames,dvipsnames]{xcolor}

\usepackage{hyperref}
\hypersetup{
  colorlinks,
  allcolors=NavyBlue,
  linktoc=all
}

\usepackage[commentmarkup=uwave]{changes}

\newcommand*\dagg{^{\dagger}}
\newcommand*{\unit}[1]{\ensuremath{\,\mathrm{#1}}}
\newcommand*{\s}[1]{\ensuremath{_\mathrm{#1}}}	% roman math subscript
\newcommand*\C{\mathcal C}

\newcommand*\nth{\bar n\s{th}}
\newcommand*\neff{n\s{eff}}

\newcommand*\Ain{\hat A\s{in}}
\newcommand*\Aout{\hat A\s{out}}

\begin{document}

\title{Optomechanical generation of a mechanical catlike state by phonon subtraction}

\author{Itay Shomroni}\email{itay.shomroni@epfl.ch}
\author{Liu Qiu}
\affiliation{Institute of Physics, \'Ecole Polytechnique F\'ed\'erale de Lausanne, CH-1015 Lausanne, Switzerland}
\author{Tobias J. Kippenberg}
\affiliation{Institute of Physics, \'Ecole Polytechnique F\'ed\'erale de Lausanne, CH-1015 Lausanne, Switzerland}

\date{23 September 2019}

\begin{abstract}
We propose a scheme to prepare a macroscopic mechanical oscillator in a cat-like state, close to a coherent state superposition.
The mechanical oscillator, coupled by radiation-pressure interaction to a field in an optical cavity, is first prepared close to a squeezed vacuum state using a reservoir engineering technique.
The system is then probed using a short optical pulse tuned to the lower motional sideband of the cavity resonance, realizing a photon-phonon swap interaction.
A photon number measurement of the photons emerging from the cavity then conditions a phonon-subtracted cat-like state with
a negative Wigner distribution
exhibiting separated peaks and multiple interference fringes.
We show that this scheme is feasible using state-of-the-art photonic crystal optomechanical system.
\end{abstract}

\pacs{Valid PACS appear here}% PACS
%\keywords{Suggested keywords}%Use showkeys class option if keyword display desired

\maketitle

\emph{Introduction.---}%
Since its inception, major questions in quantum mechanics have been whether and how the superposition principle applies to macroscopic objects, as embodied in the famous thought experiment of Schr\"odinger.
Nowadays,
superpositions of coherent states, also known as cat states,
are routinely generated in microscopic systems such as ions~\cite{monroe1996,leibfried2005}, radiation in superconducting cavities~\cite{vlastakis2013}, optical photons~\cite{huang2015,ourjoumtsev2006,neergaard2006}, as well as atoms~\cite{omran2019} and hybrid atom-light systems~\cite{hacker2019}.
Preparing such states in macroscopic systems has proved to be more difficult.
It has mainly been considered within the framework of quantum optomechanics, where a macroscopic mechanical oscillator is coupled to an electromagnetic field in a cavity via radiation pressure interaction~\cite{aspelmeyer2014}.
Recent advances in optomechanics include cooling the oscillator to its ground state~\cite{chan2011,qiu2019},
observation of quantum correlations between light and mechanics~\cite{purdy2013b,purdy2013,safavi-naeini2013,sudhir2017,safavi-naeini2012,sudhir2017}, and quantum nondemolition (QND) measurements~\cite{suh2014,shomroni2019} and squeezing~\cite{kronwald2013,wollman2015,lecocq2015,pirkkalainen2015,lei2016} of mechanical motion,
conditional preparation of single-phonon and entangled mechanical states~\cite{galland2014,hong2017,riedinger2018,marinkovic2018},
continuous-variable steady-state mechanical entanglement~\cite{ockeloen-korppi2018},
and generation of acoustic Fock states through coupling with superconducting qubits~\cite{chu2017,chu2018,satzinger2018,sletten2019,arrangoiz-arriola2019}.
Despite numerous theoretical proposals~\cite{mancini1997,bose1997,kleckner2008,paternostro2011,romero-isart2011,pepper2012,akram2013,tan2013,vanner2013,milburn2016,hoff2016,liao2016,abdi2016,clarke2018,li2018,davis2018},
however, the generation of macroscopic 
%continuous-variable
cat states has remained elusive.
Such states are interesting not only for addressing fundamental questions in quantum theory~\cite{marshall2003,blencowe2013,nimmrichter2014,carlesso2019}, but also for efficient encoding of quantum information in continuous-variable systems~\cite{cochrane1999,leghtas2013,mirrahimi2014}

\begin{figure}
	\includegraphics[scale=1]{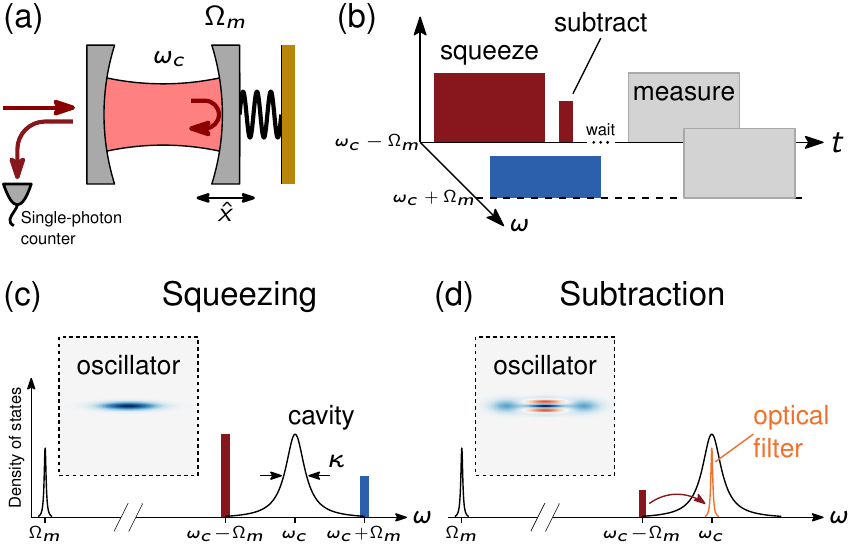}
	\caption{Optomechanical scheme for generation of a mechanical catlike state.
	(a)~Illustration of a cavity-optomechanical system. A mechanical oscillator (frequency $\Omega_m$, energy dissipation rate $\Gamma_m$, displacement $\hat x$) forms part of an optical cavity (frequency $\omega_c$, energy dissipation rate $\kappa$).
	Cavity photons couple to the oscillator through radiation-pressure interaction, and the output light from the cavity is analyzed.
	(b)~Time-domain picture of the scheme.
	The oscillator is first prepared in a squeezed state by driving the cavity on the upper and lower motional sidebands.
	Then, a short pulse on the lower motional sideband drives an anti-Stokes photon-phonon scattering (beamsplitter interaction), subtracting phonons from the mechanical state, which can be analyzed after a variable wait time.
	(c,d) Frequency-domain pictures of the squeezing (c) and subtraction (d) stages.
	The Wigner distribution of the mechanical state is also shown.
	}
	\label{fig:scheme}
\end{figure}

Here we propose a scheme for preparation of a mechanical catlike state in quantum optomechanics that does not rely on nonlinear coupling~\cite{romero-isart2011,tan2013,akram2013} or on external generation and transfer of the nonclassical state~\cite{hoff2016,teh2018}.
An established technique for preparation of cat-like states is subtraction of one or several quanta from a squeezed vacuum state~\cite{dakna1997,biswas2007}.
This was demonstrated in optics by passing a squeezed vacuum beam through a high-transmission beam splitter.
Conditioned on $m$ photons detected at the reflection port, an equal number of photons is subtracted from the transmitted beam, resulting in a catlike state~\cite{ourjoumtsev2006,neergaard2006}.
In optomechanics, it was proposed to transfer such 
a state
%photon-subtracted squeezed vacuum 
onto a mechanical oscillator~\cite{hoff2016}.
Ref.~\onlinecite{milburn2016} studied nonclassical state generation through various combinations of pulsed position measurements or measurement-induced mechanical squeezing, with single-phonon subtraction or addition, using two optical modes.

Our approach
%combines cavity optomechanical quantum control and number-resolved photon counting using a single cavity mode in the resolved-sideband regime, 
%
combines methods adopted from continuous quantum control with number-resolved photon counting, both available in a single sideband-resolved optical mode,
to 
%directly
generate a phonon-subtracted squeezed mechanical state (Fig.~\ref{fig:scheme}).
In optomechanics, it is possible to realize a beamsplitter-type interaction between a cavity photon (frequency $\omega_c$) and a mechanical excitation (a phonon of frequency $\Omega_m$) within the resolved-sideband regime $\Omega_m\gg\kappa$, where $\kappa$ is the cavity linewidth~\cite{aspelmeyer2014}.
This interaction occurs when the cavity is driven with frequency $\omega_l$ on the lower motional sideband, $\omega_c-\omega_l=-\Omega_m$, through cavity-enhanced anti-Stokes scattering of drive photons by the oscillator
(Fig.~\ref{fig:scheme}d), and
is the basis of sideband cooling of mechanical motion~\cite{aspelmeyer2014} and coherent photon-phonon swap~\cite{hofer2011,palomaki2013}.
In our scheme, shown in Fig.~\ref{fig:scheme},
the mechanical oscillator is first prepared close to a squeezed vacuum state~\cite{kronwald2013,wollman2015,lecocq2015,pirkkalainen2015,lei2016},
and then one or several phonons are swapped with photons which proceed to emerge from the cavity.
Light from the cavity is optically filtered on the resonance frequency in order to detect only anti-Stokes scattered photons.
Conditioned on subsequent number-resolved photon detection~\cite{mattioli2015}, a mechanical phonon-subtracted squeezed state is generated.
The state can be subsequently analyzed by mechanical tomography, for example using single-quadrature QND measurements, demonstrated in the optical domain~\cite{shomroni2019}, or by state swap followed by homodyne detection~\cite{palomaki2013}.

Optomechanical crystals~\cite{chan2012} are an especially promising platform for our scheme.
They operate in the resolved sideband regime, and 
%high-fidelity 
cooling to the ground state with strong driving has been demonstrated~\cite{qiu2019}
(note that cooling and squeezing are here combined in a single step~\cite{kronwald2013}).
Additionally, they can show extremely long coherence times of more than a second~\cite{maccabe2019}, making them attractive for studying nonclassical states of motion.
We note that the individual components of our scheme have both been separately implemented.
Squeezing was successfully demonstrated in several optomechanical systems~\cite{wollman2015,lecocq2015,pirkkalainen2015,lei2016}, and photon counting was applied to optomechanical crystals prepared in the ground state to generate single-phonon and entangled~\cite{cohen2015,riedinger2016,hong2017,riedinger2018,marinkovic2018} mechanical states.

\begin{figure}
	\includegraphics[scale=1]{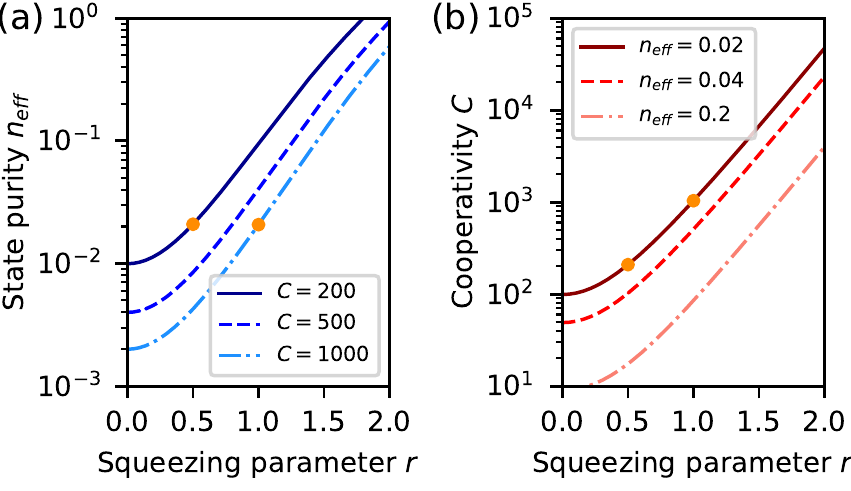}
	\caption{State purity in optomechanical dissipative squeezing.
	(a)~State purity~$\neff$ vs.~squeezing parameter~$r$ for different cooperativities~$\C$.
	(b)~The cooperativity required to achieve a given purity vs.~$r$.
	The steady-state in optomechanical dissipative squeezing, in particular the variance of the squeezed quadrature but also the thermal component $\neff$, results from a trade-off between optical damping and ratio of the drives.
	The two working points with $\neff=0.02$ used in this work are indicated in both panels.
	}
	\label{fig:purity}
\end{figure}

\emph{Squeezing of the mechanical state.---}%
The first stage of our protocol is squeezing of the mechanical oscillator by reservoir engineering~\cite{kronwald2013}.
The optomechanical system in the resolved-sideband regime is driven with two tones tuned to the upper ($+$) and lower ($-$) mechanical sidebands, with coupling rates $g_\pm = g_0\sqrt{\bar n_\pm}$
(Fig.~\ref{fig:scheme}c),
where $g_0$ is the single-photon optomechanical coupling rate and $\bar n_\pm$ is the mean intracavity photon number due to each drive.
When $g_+=g_-$, a QND measurement of a single mechanical quadrature $\hat X_1 = (\hat b\dagg+\hat b)/\sqrt{2}$ is performed~\cite{caves1980,clerk2008}, with $\hat b$ being the phonon annihilation operator.
When $g_->g_+$, however, both quadratures $\hat X_1$ and $\hat X_2=i(\hat b\dagg-\hat b)/\sqrt{2}$ are equally damped by the cavity field while the fluctuations associated with the damping are distributed unequally.
This results in a squeezed thermal state characterized by a squeezing parameter $r$ and purity $\neff$, where $\tanh r = g_+/g_-$ and
$\neff+\frac{1}{2}=\sqrt{\langle\Delta X_1^2\rangle\langle\Delta X_2^2\rangle}$.
The advantage of this scheme is that it allows arbitrarily strong squeezing (limited by drive power), in particular exceeding the $3\unit{dB}$ limit of parametric driving.
While Ref.~\onlinecite{kronwald2013} focused on maximum squeezing (minimum variance in one quadrature) for a given drive power characterized by the cooperativity~$\C=4g_-^2/\kappa\gamma$,
this comes at the expense of increased $\neff$ (although for optimal squeezing, $\neff\rightarrow 0.2$ in the limit of high cooperativity~\cite{kronwald2013}).
State purity, however, is important for engineering quantum states, and in this work we relax the demand for optimal squeezing in favor of purity.
For a given cooling strength $\C$, there is a trade-off between the state purity $\neff$ and the amount of squeezing $r$ related to the imbalance of the drives~\cite{kronwald2013}.
Figure~\ref{fig:purity}a shows the state purity $\neff$ vs.~the squeezing parameter $r$ for different cooperativities, and Fig.~\ref{fig:purity}b shows the required cooperativity vs.~the squeezing parameter for different purities.
In Fig.~\ref{fig:purity}, we assume that the mechanical oscillator is coupled to a bath with mean thermal occupancy $\nth=2$.
For conciseness, in this work we consider two working points, both with $\neff=0.02$:
(1) $r=0.5$ ($4.3\unit{dB}$ squeezing), $\C\simeq 200$, compatible with recent 
high fidelity
ground state cooling experiments in optomechanical crystals~\cite{qiu2019} and
(2) $r=1$ ($8.7\unit{dB}$), $\C\simeq 1000$.
Note that mechanical squeezing of $4.7\unit{dB}$ has been reported~\cite{lei2016}.
Accordingly, we will assume in the following that the oscillator is prepared in the desired squeezed state.

\emph{Conditional phonon subtraction.---}%
Following the squeezing stage, we apply a weak pulse tuned to the lower motional sideband (Fig.~\ref{fig:scheme}d), realizing a beamsplitter interaction,
$\hat H\s{int} = \hbar g(\hat a\dagg\hat b+\hat b\dagg\hat a)$,
where $\hat a$ is the photon annihilation operator in a frame displaced by the mean cavity field $\bar a$, and $g=g_0\bar a$ is the coupling rate enhanced by $\bar a $~\cite{suppmat}.
The relation between the mechanical mode $\hat b(t)$ and the cavity output field $\Aout(t)$ assuming weak coupling $g\ll\kappa$ is (see the appendix)
\begin{equation}
\begin{pmatrix}\Aout(t) \\ \hat b(t) \end{pmatrix} =
\begin{pmatrix}\cos\theta & i\sin\theta \\ i\sin\theta & \cos\theta \end{pmatrix}
\begin{pmatrix}\Ain(t)\\ \hat b(0) \end{pmatrix}
\label{eq:beamsplitter1}
\end{equation}
where $\cos\theta\equiv e^{-\tilde g t}$ is the beamsplitter amplitude ``transmission'' with $\tilde g\equiv 2g^2/\kappa$ being the interaction strength and $\Ain(t)$ being the optical input in the second ``port'' of the beamsplitter, which is vacuum noise in the displaced frame.
If the initial mechanical state is squeezed vacuum ($\neff=0$)
$\hat S(r)\hat\rho_0\hat S\dagg(r)$ where $\hat\rho_0 = \lvert 0\rangle\langle 0\rvert$
and $\hat S(r) = e^{r(\hat b^2-\hat b\dagg{}^2)/2}$ is the squeezing operator, 
the final mechanical state $\hat\rho\s{out}^{(m)}$ conditioned on the detection of $m$ photons in the output field can be calculated analytically~\cite{dakna1997}.
It is parametrized by $m$ and by $\cos^2\theta\tanh r$, 
with the initial squeezing degraded by the transmission $\cos^2\theta$
due to mixing with the optical vacuum noise $\Ain$.
Increasing the transmission however also reduces the probability to herald $m$ subtracted phonons.
Importantly, as we discuss below, non-negligible values of $\theta$ can be easily obtained with pulse durations satisfying $\kappa^{-1} \ll t\s{pulse} \ll (\nth\Gamma_m)^{-1}$, with $\nth\Gamma_m$  being the thermal decoherence rate;
thus, we can neglect decoherence of the mechanical state during the pulse.
The Wigner distribution of the conditioned mechanical state appears as two displaced peaks with an intermediate oscillating region, similar to a cat state.
The peak separation increases with $m$ and initial squeezing~$r$~\cite{dakna1997}.

\begin{figure}
	\includegraphics[scale=1]{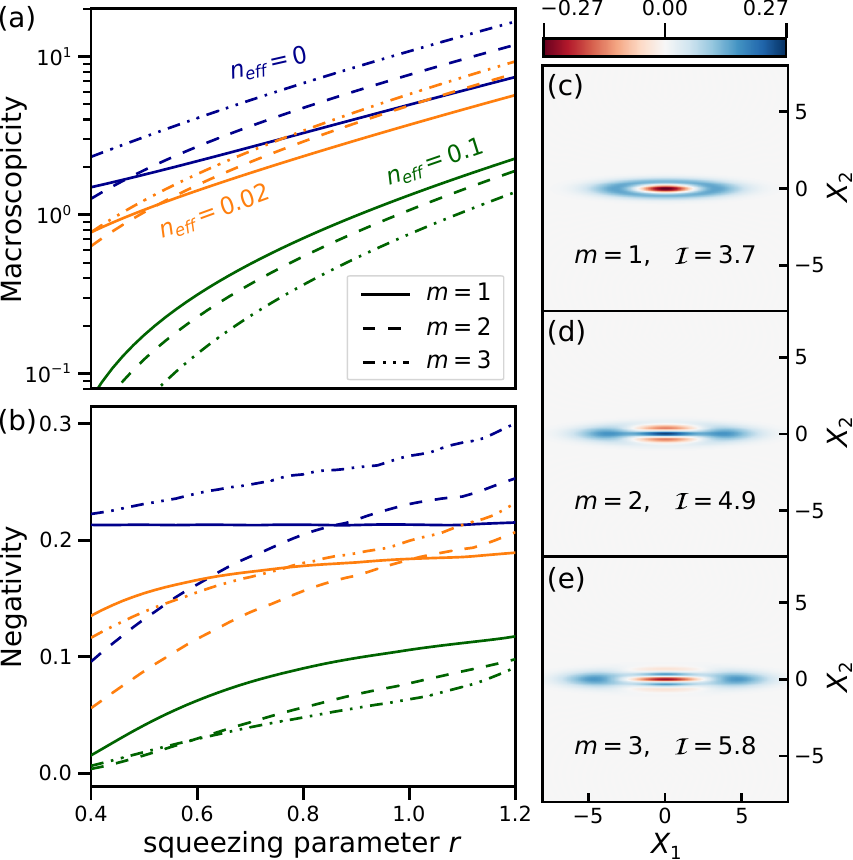}
	\caption{Effect of initial state purity on the nonclassicality of phonon-subtracted squeezed thermal mechanical states.
	(a)--(b)~The output state macroscopicity and Wigner negativity vs.~squeezing parameter $r$ for state purities $\neff=0$ (blue), 0.02 (orange), and 0.1 (green) and for~$m=1$ (solid), 2~(dashed), and 3~(dash-dotted) subtracted phonons.
	Squeezed vacuum $\neff=0$ is shown for reference.
	(c)--(e) Wigner distributions for $r=1$, $\neff=0.02$, and $m=1,2,3$.
	The macroscopicity $\mathcal{I}$ is indicated.
	In all panels, we set $\theta=0.1$.}
	\label{fig:negativity}
\end{figure}

Squeezed Fock and and thermal states have also been treated in the literature~\cite{kim1989,marian1991,marian1992,marian1993,hu2010} but yield complicated expressions.
Instead we solve numerically for the mechanical output state when the input is a squeezed thermal state, with parameters $m$, $r$, $\neff$, and $\theta$~\cite{suppmat}.
As in Refs.~\mbox{\onlinecite{hoff2016,clarke2018}}, we 
characterize the quantum nature of the output state using two measures based on the Wigner distribution $W(x,p)$.
The macroscopicity~\cite{lee2011}
%\footnote{The macroscopicity of a single-mode state $\rho$ can be conveniently calculated as $\mathcal{I}(\rho)=-\mathrm{tr}(\rho\mathcal{L}\rho)$ where $\mathcal{L}\rho = a\rho a\dagg -\frac{1}{2}\rho a\dagg a-\frac{1}{2}a\dagg a\rho$.}
\begin{equation}
\mathcal{I} = -\frac{\pi}{2}\int\!\!\!\int W(x,p)\biggl(\frac{\partial^2}{\partial x}+\frac{\partial^2}{\partial p}+2\biggr)W(x,p)\,dx\,dp
\end{equation}
assesses $W(x,p)$ through the amplitude and frequency of its interference fringes,
with higher values indicating higher nonclassicality.
For any state with a given mean excitation number $\langle\hat n\rangle$, the maximum possible value of $\mathcal{I}$ is $\langle\hat n\rangle$.
In particular, this maximum is attained both for cat states and for phonon-subtracted squeezed vacuum states, but also for squeezed vacuum~\footnote{$\mathcal{I}$ is maximized for any pure state that is orthogonal to a phonon-subtracted state of itself~\cite{lee2011}.
Cat states, phonon-subtracted squeezed vacuum states, and squeezed vacuum, all contain only odd or only even number states, and thus satisfy this condition.}.
We also consider the Wigner negativity~\cite{kenfack2004}
\begin{equation}
\mathcal{N} = \frac{1}{2}\biggl(\int\!\!\!\int\lvert W(x,p)\rvert\,dx\,dp-1\biggr)
\end{equation}
which is simply the phase-space volume of the negative part of $W(x,p)$.
Figure~\ref{fig:negativity}a,b shows these measures vs.~the squeezing parameter~$r$ for different initial mechanical state purities~$\neff$ and for detection of~$m=1, 2$ or~3 photons.
As expected, the nonclassicality of the final mechanical state is degraded by initial state impurity, but can be increased by stronger squeezing.
Figure~\ref{fig:negativity}a,b also shows that for highly impure initial states or very weak squeezing, more subtractions actually decrease nonclassicality.

\begin{figure*}
	\includegraphics[scale=1]{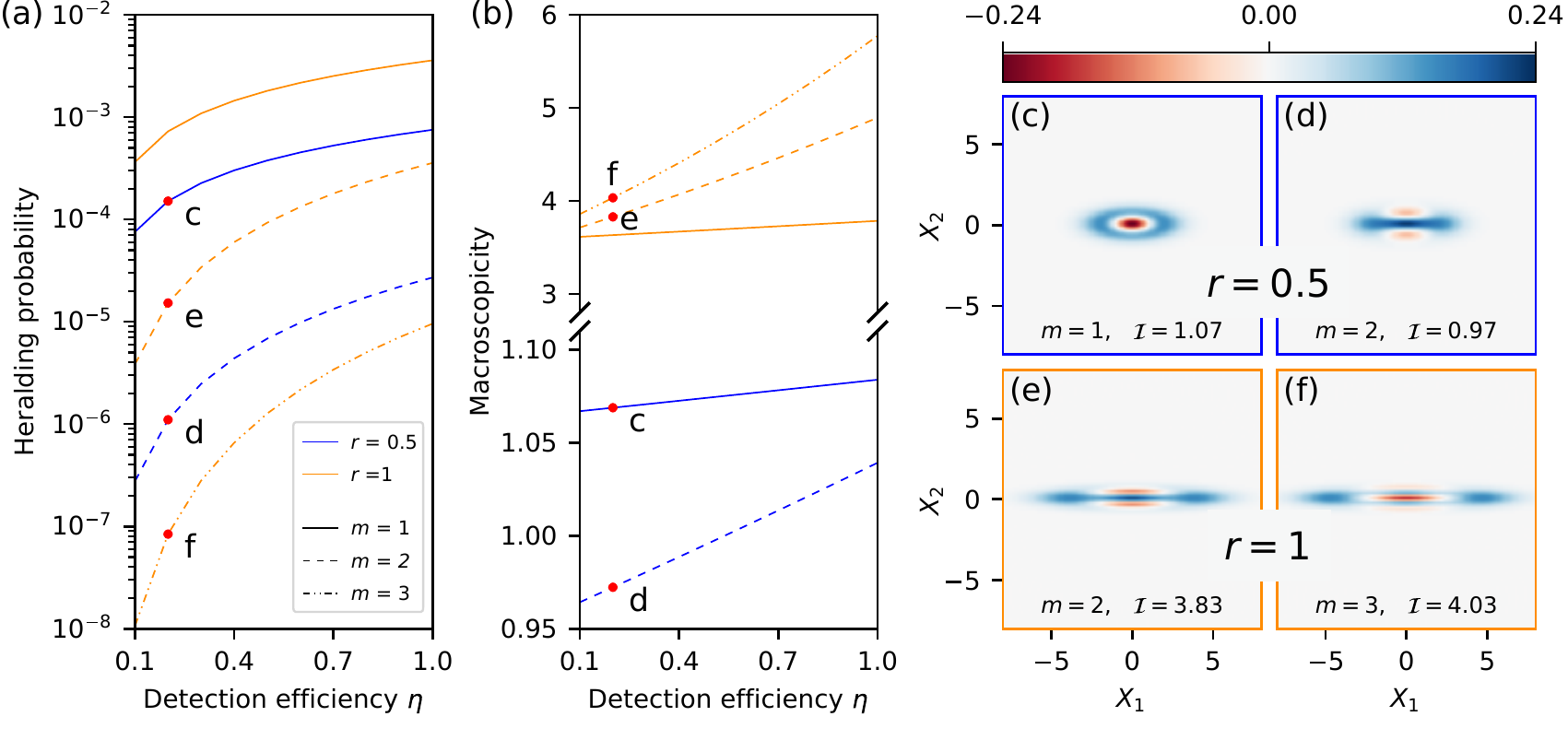}
	\caption{Effect of optical losses on generated catlike mechanical state.
	(a)~Heralding probability (successful detection of $m$ photons) vs.~total optical detection efficiency $\eta$, shown for initial squeezing $r=0.5$ (blue) and $r=1$ (orange), and for $m=1$~(solid), $2$~(dashed), and $3$~(dash-dotted) subtracted phonons.
	This probability incorporates the weak optomechanical beam splitter interaction $\theta$, chosen as $\theta=0.05$ for $m=1$ cases and $\theta=0.1$ for $m=2$ cases.
	(b)~The macroscopicity $\mathcal{I}$ corresponding to the same curves in panel~(a).
	(c--f)~Wigner distributions and macroscopicities of the heralded mechanical state, assuming $\eta=0.2$, for the points indicated in panels (a) and~(b).}
	\label{fig:losses}
\end{figure*}

Figure~\ref{fig:negativity}c--e shows the Wigner distributions for $r=1$, $\neff=0.02$, and $m=1,2,3$, indicating the achieved macroscopicity $\mathcal{I}$.
Figure~\ref{fig:negativity} assumes no additional optical losses and thus gives maximum nonclassicality for the given parameters.
Even with optical losses, this maximum can be maintained by reducing the interaction strength at the expense of heralding probability, such that any photon lost will prevent heralding.
In an actual experiment, a balance must be struck between constraints on experiment duration and decoherence of the mechanical state due to optical losses.
We extend the previous analysis by including a beam splitter in the optical path
to account for a finite optical detection efficiency~$\eta$.
Figure~\ref{fig:losses}a,b shows the effect on the heralding probability and macroscopicity, respectively, in the case $\neff=0.02$ for $r=0.5,1$ and $m=1,2,3$.

\emph{Experimental realization.---}%
To estimate the experimental feasibility of our scheme we consider an optomechanical  crystal, similar to that used in our recent QND~\cite{shomroni2019} and ground-state cooling~\cite{qiu2019} experiments, operating in the resolved sideband regime, with mechanical frequency $\Omega_m/2\pi=5.2\unit{GHz}$, intrinsic mechanical linewidth $\Gamma_m/2\pi=100\unit{kHz}$, and optical cavity at telecommunication wavelengths with linewidth $\kappa/2\pi=1\unit{GHz}$ of which $\kappa\s{ex}/2\pi=800\unit{MHz}$ output coupling and the rest intrinsic dissipation~\footnote{In other words, we use an overcoupled version of the system of Ref.~\onlinecite{qiu2019} with $\kappa_i/2\pi\simeq 200\unit{MHz}$ intrinsic cavity energy loss rate.}.
We assume cryogenic operation at temperature $T\s{bath}=0.5\unit{K}$, yielding bath occupation $\nth=2$ and thermal decoherence time of $(n\s{th}\Gamma_m)^{-1}\approx 1\unit{\mu s}$, much longer than the cavity lifetime and mechanical period, and thus we can safely neglect thermal decoherence in the analysis.
Note that optomechanical crystals with decoherence times above $1\unit{s}$ have been demonstrated, albeit at mK temperatures~\cite{maccabe2019}.
With a typical single-photon optomechanical coupling rate $g_0/2\pi\sim 1\unit{MHz}$, beamsplitter reflections of the order of a few percent as used here can be realized using, e.g., $10\unit{ns}$ pulses of low input power $\sim 10\unit{\mu W}$, much weaker than in typical cooling experiments.

We assume total detection efficiency $\eta=20\%$, 
which is feasible assuming almost 100\% outcoupling of light from the cavity into an optical fiber, as was demonstrated in Refs.~\onlinecite{tiecke2015,burek2017,magrini2018}; 80\% cavity efficiency $\kappa_e/\kappa$; and additional 25\% efficiency due to the transmission of the optical filter and other components, and photon-counter detection efficiency.
This gives heralding probabilities in the range $10^{-4}$--$10^{-7}$ from Fig.~\ref{fig:losses}a.
Squeezing of the initial thermal state occurs at a rate $\C\Gamma_m\approx 2\pi\times 20\unit{MHz}$ for $\C=200$, and thus reducing the initial occupancy $\nth=2$ to $\neff=0.02$ can be done in a timescale of $\sim 100\unit{ns}$.
As noted above, the subtraction pulse duration can be $\sim 10\unit{ns}$.
We assume next a tomography of the final mechanical state to take $\sim 100\unit{ns}$, given the mechanical period of $\sim 30\unit{ps}$.
Overall we conservatively assume a repetition rate $\sim 10\unit{\mu s}$.
Thus even with a heralding probability of $10^{-7}$, we expect 1 event every $100\unit{s}$, resulting in a feasible experiment duration of several hours.
Note that similar photon-counting experiments were done on a time scale of $100\unit{hrs}$~\cite{hong2017}.

Figure~\ref{fig:losses}c--f shows the mechanical Wigner distributions corresponding to $\eta=0.2$.
For squeezing $r=0.5$, macroscopicities $\mathcal{I}\approx 1$ are obtained, similar to a single phonon Fock state but with substantially different distributions (Fig.~\ref{fig:losses}c,d).
For $r=1$ much higher values $\mathcal{I}\approx 4$ are possible (Fig.~\ref{fig:losses}e,f).

\emph{Conclusion.---}%
We presented a scheme to prepare a macroscopic mechanical oscillator in a cat-like state by combining reservoir-engineering techniques, phonon-photon swap operations, and photon counting.
A key feature of our scheme is its simplicity.
It does not require preparation of nonclassical states of light, and is similar to methods used to generate macroscopic Fock states~\cite{galland2014,hong2017}, differing essentially in the squeezing step.
We have used experimental parameters that 
are currently available in optomechanical crystals.
While in this work we considered phonon subtraction from a squeezed state, phonon addition may equally well be performed, by applying a pulse tuned to the \textit{upper} motional sideband, providing additional avenues for generating nonclassical mechanical states~\cite{milburn2016,li2018}.
Generation of such states will enable the study of quantum theory in macroscopic objects, and is a first step in using highly coherent and scalable mechanical platforms for continuous variable quantum information applications~\cite{cochrane1999}.

\begin{acknowledgments}
	We thank D.~Malz, C.~Galland, and N.~J.~Engelsen for useful discussions and comments.
	This work was supported by the European Union's Horizon 2020 research and innovation programme under Grant Agreement No.~732894 (FET Proactive HOT).
\end{acknowledgments}

\appendix*
\section{Theoretical details}

We consider a standard optomechanical system with a single relevant optical mode with frequency $\omega_c$, and annihilation operator $\hat a$ and a single relevant mechanical mode with frequency $\Omega_m$ and annihilation operator $\hat b$.
The two modes interact via radiation pressure interaction with a single-photon coupling rate $g_0$.
The Hamiltonian is given by~\cite{aspelmeyer2014}
\begin{multline}
\hat H = \hbar\omega_c\hat a\dagg\hat a 
       + \hbar\Omega_m\hat b\dagg\hat b 
       - \hbar g_0\hat a\dagg\hat a(\hat b\dagg+\hat b) \\
       -i\hbar[\hat a\s{in}(t)\hat a\dagg - \hat a\s{in}(t)\dagg\hat a],
\label{eq:Hamiltonian0}
\end{multline}
where $\hat a\s{in}(t) = e^{-i\omega_l t}(\bar a\s{in} + \delta\hat a\s{in})$
is a coherent drive at frequency $\omega_l$, separated into its mean and noise terms.
For simplicity, we take the drive envelope to be constant.
Following standard procedure, we move to a frame rotating with the drive frequency and linearize the dynamics about the mean values
$\hat a = \bar a + \delta\hat a$ and $\hat b = \bar b + \delta\hat b$.
This yields 
\begin{multline}
H = -\hbar\Delta\delta\hat a\dagg\delta\hat a 
  + \hbar\Omega_m\delta\hat b\dagg\delta\hat b 
  - \hbar g(\delta\hat a+\delta\hat a\dagg)(\delta\hat b\dagg+\delta\hat b) \\
-i\hbar\bar a\s{in}(\delta\hat a\dagg - \delta\hat a),
\label{eq:Hamiltonian1}
\end{multline}
where $\Delta=\omega_l-\omega_c$ and $g = g_0\bar a$.
Including dissipation to the optical and mechanical baths as well as the adjoining input noises yields the Langevin equations
\begin{subequations}
\begin{align}
\dot{\hat a} &= -(\kappa/2-i\Delta)\hat a - ig(\hat b\dagg +\hat b) + \sqrt{\kappa}\hat a\s{in},\\
\dot{\hat b} &= -(\Gamma_m/2+i\Omega_m)\hat b - ig(\hat a\dagg + \hat a) + \sqrt{\Gamma_m}\hat b\s{in},
\end{align}
\end{subequations}
where for brevity we have omitted the $\delta$ designation on the noise operators.
When the drive is tuned to the lower mechanical sideband, $\Delta=-\Omega_m$, and in the good cavity limit, $\kappa\ll\Omega_m$, we can perform the rotating-wave approximation, leading to
\begin{subequations}
\begin{align}
\dot{\hat a} &= -(\kappa/2-i\Delta)\hat a + ig\hat b + \sqrt{\kappa}\hat a\s{in},\\
\dot{\hat b} &= -(\Gamma_m/2+i\Omega_m)\hat b + ig\hat a + \sqrt{\Gamma_m}\hat b\s{in},
\end{align}
\end{subequations}

We will be interested in interactions occurring on a time scale much shorter than the mechanical dissipation, and hence we can neglect $\Gamma_m$. We further move into a frame rotating at the mechanical frequency to obtain
\begin{subequations}
\begin{align}
\dot{\hat a} &= -\frac{\kappa}{2}\hat a + ig\hat b + \sqrt{\kappa}\hat a\s{in},\\
%\dot{\hat b} &= -\frac{\Gamma_m}{2}\hat b + ig\hat a + \sqrt{\Gamma_m}\hat b\s{in}
\dot{\hat b} &=  ig\hat a,
\end{align}
\end{subequations}
In the weak coupling limit $g\ll\kappa$, we can adiabatically eliminate the cavity dynamics,
\begin{subequations}
	\begin{align}
	\hat a(t) &= i\frac{2g}{\kappa}\hat b + \frac{2}{\sqrt{\kappa}}\hat a\s{in},
	\label{eq:A6a}\\
	\hat b(t) &= e^{-\tilde g t}\hat b(0) + i\sqrt{2\tilde g}\,e^{-\tilde g t}\int_0^t dt' e^{\tilde g t'}\hat a\s{in}(t'),
	\label{eq:A6b}
	\end{align}
\end{subequations}
where we defined the coupling strength $\tilde g = 2g^2/\kappa$.
Substituting the output field given by $\hat a\s{out} = -\hat a\s{in}+\sqrt{\kappa}\hat a$ in Eq.~\eqref{eq:A6a} yields
%\begin{subequations}
%	\begin{align}
%	\hat a\s{out} &= \hat a\s{in} + i\sqrt{2\tilde g}\hat b\\
%	\dot{\hat b} &=  -\tilde g\hat b + i\sqrt{2\tilde g}\hat a\s{in}
%	\label{eq:bdot}
%	\end{align}
%	\label{eq:A7}
%\end{subequations}
\begin{equation}
\hat a\s{out} = \hat a\s{in} + i\sqrt{2\tilde g}\hat b.
\label{eq:A7}
\end{equation}

We introduce the temporal optical modes~\cite{hofer2011,galland2014}
\begin{subequations}
	\begin{align}
	\Ain(t)  &= \sqrt{\frac{2\tilde g}{e^{2\tilde g t}-1}}
				\int_0^t dt' e^{\tilde g t'}\hat a\s{in}(t'), \\ 
	\Aout(t) &= \sqrt{\frac{2\tilde g}{1-e^{-2\tilde g t}}}
				\int_0^t dt' e^{-\tilde g t'}\hat a\s{out}(t'),
	\end{align}
\end{subequations}
which obey $[\hat A_i,\hat A_i\dagg]=1$, in Eqs.~\eqref{eq:A6b} and~\eqref{eq:A7} to yield
\begin{subequations}
	\begin{align}
	\Aout(t) &= e^{-\tilde g t}\Ain(t) + i\sqrt{1-e^{-2\tilde g t}}\,\hat b(0),\\
	\hat b(t) &= e^{-\tilde g t}\hat b(0) + i\sqrt{1-e^{-2\tilde g t}}\,\Ain(t).
	\end{align}
\end{subequations}
In other words, we realize a beam-splitter transformation between mechanical and optical modes
\begin{equation}
\begin{pmatrix}
\Aout(t) \\
\hat b(t)
\end{pmatrix}
=
\begin{pmatrix}
\cos\theta & i\sin\theta \\
i\sin\theta & \cos\theta
\end{pmatrix}
\begin{pmatrix}
\Ain(t)\\
\hat b(0)
\end{pmatrix}
\label{eq:beamsplitter}
\end{equation}
with $\cos\theta\equiv e^{-\tilde g t}$ and $\sin\theta\equiv\sqrt{1-e^{-2\tilde g t}}$.
In our case, $\tilde gt\ll 1$ so $\theta\ll 1$.
The unitary transformation~\eqref{eq:beamsplitter} entails~\cite{dakna1997,campos1989}
$\Aout = U\dagg\Ain U$ and $\hat b(t) = U\dagg\hat b(0)U$ with $U=e^{i\frac{\pi}{2}\hat L_3}e^{-2i\theta\hat L_2}e^{-i\frac{\pi}{2}\hat L_3}$
and the ``angular momentum'' operators given by~\cite{yurke1986,ban1994,dakna1997}
\begin{subequations}
	\begin{align}
	\hat L_2 &= \frac{1}{2i}[\Ain\dagg(t)\hat b(0)-\hat b\dagg(0)\Ain(t)],\\
	\hat L_3 &= \frac{1}{2}[\Ain\dagg(t)\Ain(t)-\hat b\dagg(0)\hat b(0)].
	\end{align}
\end{subequations}

Thus, in the Schr\"odinger picture, a system initially described by a density matrix
$\hat\rho\s{in}$ will evolve according to $\hat\rho\s{out}=\hat U\hat\rho\s{in}\hat U\dagg$.
The initial state of the systems is 
\begin{equation}
\hat\rho\s{in} = \hat\rho\s{in}^M\otimes\lvert 0\rangle\langle 0\rvert,
\end{equation}
where $\hat\rho\s{in}^M$ is the mechanical input state and the cavity is in the vacuum state (hereafter all bras and kets refer to optical Hilbert space).
In this case the output state is given by~\cite{ban1994,dakna1997}
\begin{equation}
\label{eq:outputstate}
\begin{split}
\hat\rho\s{out} &= \sum_{n=0}^{\infty}\sum_{m=0}^{\infty} \biggl[
	\frac{e^{-i(m-n)\pi/2}}{\sqrt{n!\,m!}}(-1)^{m+n}\lvert\tan\theta\rvert^{m+n} \\
	&\quad\times \hat b^m \lvert\cos\theta\rvert^{\hat b\dagg\hat b}
		   \hat\rho\s{in}^M
		   \lvert\cos\theta\rvert^{\hat b\dagg\hat b} \hat b\dagg{}^n
    \otimes \lvert m\rangle\langle n\rvert\biggr].
\end{split}
\end{equation}
Conditioned on detection of $m$ output photons, the mechanical state is reduced in the usual way
\begin{equation}
\label{eq:reducedstate}
\rho\s{out}^{(m)} = 
	\frac{\langle m\rvert\hat\rho\s{out}\rvert m\rangle}{\mathrm{tr}_M(\langle m\rvert\hat\rho\s{out}\rvert m\rangle)}
\end{equation}
with probability
\begin{equation}
\label{eq:prbability}
\begin{split}
P(m) &= \mathrm{tr}_M(\langle m\rvert\hat\rho\s{out}\rvert m\rangle) \\
	 &= \sum_{n=m}^{\infty} \binom{n}{m}(\sin\theta)^{2m}(\cos\theta)^{2(n-m)}
	    \langle n\vert\hat\rho\s{in}^M\vert n\rangle.
\end{split}
\end{equation}
Equations~\eqref{eq:outputstate}, \eqref{eq:reducedstate} and~\eqref{eq:prbability} can be solved for an arbitrary mechanical input state $\hat\rho\s{in}^M$.
This has been done for squeezed vacuum in Ref.~\onlinecite{dakna1997}.
In our work we solve numerically for $\hat\rho\s{out}^{(m)}$ for a squeezed thermal input state.

%\clearpage

\bibliography{refs}

\end{document}